\documentclass{article}
\usepackage{graphicx}

\newcommand{\kk}[2]{\frac{#1}{#2}}
\newcommand{\pab}[2]{\kk{\p #1}{\p #2}}
\def\be{\begin{equation}}
\def\ee{\end{equation}}
\def\p{\partial}
\def\a{\alpha}

\def\na{\nabla}
\def\g{\gamma}

\def\s{\;\;}
\begin{document}

\title{Computational Modelling of Nonlinear Calcium Waves}

\author{Xin-She Yang  \\
Department of Engineering, University of Cambridge \\
Trumpington Street, Cambridge CB2 1PZ  }

\date{}

\maketitle

\begin{abstract}
The calcium transport in biological systems is modelled as a
reaction-diffusion process. Nonlinear  calcium waves are then
simulated using a stochastic cellular automaton whose rules are
derived from the corresponding coupled partial differential
equations. Numerical simulations show self-organized criticality in
the complex calcium waves and patterns. Both the stochastic cellular
automaton approach and the equation-based simulations can predict
the characteristics of calcium waves and complex pattern formation.
The implication of locality of calcium distribution
with positional information in biological systems is also discussed. \\



\noindent {\sf Citation detail:} X. S. Yang, Computational modelling of nonlinear calcium waves, 
{\it Applied Mathematical Modelling}, {\bf 30}(2), 200-208 (2006).

\end{abstract}

\maketitle

\section{Introduction}

Many processes in living organisms such as muscle mechanics, cardiac
electrophysiology, adaptation in photo-receptors, and cytoplasm
functions involve the calcium ion transport and its physiological
functions[1-6].  However, the exact function of Ca$^{2+}$
oscillations and transport is only partially understood although it
is believed that they involve in the intracellular communications
and synchronization in the response to a local stimulus [13-15].
Even in the simplest one-dimensional case, the proper modelling
requires many simplifications and assumptions. The nonlinearity and
cross-coupling in the transport mechanisms usually lead to
intractable governing equations. Even with certain approximations
and simplicity, the mathematical models still lead to nonlinear
reaction-diffusion equations if the essence of the calcium ion
activities is included.

Nonlinear reaction-diffusion systems can exhibit complex pattern
formation [9-12]. The nonlinear system can be simulated using finite
difference or finite element methods, or the alternative cellular
automata [16]. However, existing implicit numerical solution schemes
are not always robust under general boundary conditions. Most of the
earlier work have mainly concerned the one-dimensional case with
piecewise linear models [10]. Meanwhile, the cellular automata
[17,18] have successfully modelled reaction-diffusion systems [16]
with relative stable pattern formation. This provides a possibility
of dealing with the difficult nonlinear problem of calcium
oscillation and waves using finite-state cellular automata with
rules derived from related partial differential equations.

On the other hand, self-organized criticality has been found in many
systems in nature [7,8] since its discovery by Bak [7] and his
colleagues. Since calcium oscillation and waves a very complicated
phenomenon with the characteristics of nonlinear reaction-diffusion
systems, it can be expected that regular patterns under certain
conditions. One natural question related to this is: Do the
self-organized criticality exist in the complexity of the calcium
transport? However, this is no existing literature addressing this
question in the context of calcium transport. This is partly because
the research on calcium activities and their mathematical modelling
for biological and physiological processes is still at a very
earlier stage [10,14-15].

This paper therefore first extends the existing mathematical models
for nonlinear  Ca$^{2+}$ waves. Then, a new stochastic finite-state
cellular automaton is then formulated to simulate the general
nonlinear calcium transport process and pattern formation. Based on
the reaction-diffusion equations, the stochastic CA will be
formulated. The pattern formation of calcium ions with realistic
parameters will be studied. The self-organized criticality will be
tested in the complex patterns of calcium concentration. The
positional pattern formation and its implication will briefly be
discussed.

\section{Mathematical Model}

\subsection{Governing Equations for Calcium Transport}

In many physiological processes within cells, Ca$^{2+}$ plays an
essential role in controlling cellular behavior and functioning in
the sense that calcium ions act as a signalling agent for a wide
range of cellular activities.  Calcium signaling is mediated through
oscillation in intracellular Ca$^{2+}$ concentration. Calcium ions
can bind to a vast number of proteins and enzymes, and the binding
can initiate a series of reactions that ends in the formation of a
chemical called inositol 1,4,5-trisphosphate (IP$_3$). The diffusion
of Ca$^{2+}$ and IP$_3$ through the cell cytosol can induce further
release of calcium ions from stores in the endoplasmic reticulum
(ER) through IP$_3$-sensitive channels. These channels are sensitive
to calcium itself, with fast activation for lower concentrations and
comparatively slower inhibition, thus leading to the calcium-induced
calcium release (CICR). Complex wave characteristics such as plane
waves and spiral waves have been observed in experimental studies
using {\it Xenopus} oocytes as pointed out by McKenzie and Sneyd
[19]. However, the detail mechanisms underlying these oscillation
and waves are only partially understood.

There are several mathematical models for Ca$^{2+}$ wave propagation
in the literature, and these include the important models such as
DeYoung and Keizer [2], Goldbeter {\it et al.} [20], Sneyd {\it et
al.} [15] and McKenzie and Sneyd [19]. However, different cell types
can results in different mathematical models, although the
fundamentals are quite similar. The nonlinearity in the mathematical
models can lead to complex pattern formations and spiral waves
[21-24]. The model we will use in this paper is an extended version
of a two-pool process. In the two-pool model, the calcium Ca$^{2+}$
concentration in the cytoplasm and the concentration in the
Ca$^{2+}$-sensitive pool satisfy the two-pool model [1]. Although it
can reproduce the oscillations and waves observed in Xenopus oocytes
and generating spiral waves in the higher dimensional situations.
The parameters for our simulations are based on the models developed
by Atri et al [1] and McKenzie and Sneyd [19]. The functioning of
IP$_3$ in the oscillation and waves has been reviewed in details by
Sneyd {\it et al.} [25], and th calcium signaling has been reviewed
by Falcke [26].

We use the variable $u(x,y,t)=$[Ca$^{2+}$] for the concentration
($\mu M)$ of calcium, $v(x,y,t)$ for the fraction of the IP$_3$
receptors that are active. The variable $p=$[IP$_3$] can be treated
as a bifurcation parameter for the reason discussed later. Then, the
nonlinear model equations for calcium transport [10,13-15] can be
written as \be \pab{u}{t}=D_u \na^2 u+J_{f}-J_p+J_l, \ee \be \lambda
\pab{v}{t}=D_v \na^2 v+g(u,v), \ee where $J_f$ models the flux of
calcium through the IP$_3$ receptor. $J_p$ models the amount the
Ca$^{2+}$ being pumped out of the cytoplasm back into the
endoplasmic reticulum or out through the plasma membrane. $J_l$
models the calcium leaking into the cell. We have \be J_f= k v
[\delta+\kk{(1-\delta)u}{k_1+u}] \varepsilon(p), \s J_p=\kk{ \gamma
u^d}{k_u^d+u^d}, \ee \be J_l=\a, \s g(u,v)=\kk{k_3^m}{k_3^m+u^m}-v,
\ee and \be \varepsilon (p)=\kk{p^n}{k_2^n+p^n}, \ee where $k_1,
k_2, k_3, k_u$ are constants and $\a, \beta, \delta$ are parameters.
In addition, $D_v=0$ is used in most existing models since most
models base on the assumption that Ca$^{2+}$ instantaneously
activates the IP$_3$ receptor.

As the number of IP$_3$ receptors remains approximately constant,
thus we may assume that $p$ is fixed and subsequently it can be
considered as a bifurcation parameter. In fact, some studies using
$p \in [0.24,0.2434]$ obtained very interesting results [22]. The
function $\varepsilon (p)$ is the fraction of $IP_3$ receptors that
have bound $IP_3$ and increases as $p$ increases. Then, the complete
nonlinear model equations become
 \be \pab{u}{t}=D_u \na^2 u + k w
[\delta+\kk{(1-\delta)u}{k_1+u}][\kk{p^n}{k_2^n+p^n}] -\kk{\g
u^d}{k_u^d+u^d}+\a, \ee  \be \lambda \kk{d v}{d
t}=\kk{k_3^m}{k_3^m+u^m}-v, \ee where $w$ is a time-dependent
inactivation variable, and $n=3, d=m=2$. Some typical values of the
related parameters are $k=3 \mu M s^{-1}$ (M=Mol/L is the molar
concentration), $k_1=k_3=0.7 \mu M, k_2=0.01 \mu M, k_v=1 \mu M,
k_u=0.27 \mu M, \delta=0.11, \a=0.15 \mu M s^{-1}, \lambda=0.2 s$,
$D_u=20 \mu m^2 s^{-1}, \gamma=2 \mu M s^{-1}$ [10].

By using the notation $s(u,v)=J_f-J_p+J_l$ and
$g(u,v)=k_3^m/(k_3^m+u^m)-v$, we can obtain the steady state
solution $(u_0,v_0)$. From equations (1) and (2) together with
equations (6) and (7), it is straightforward to check that they
satisfy the Turing instability conditions \be Q=(\begin{array}{cr} s_u & s_v \\
g_u & g_v \end{array}), \ee which requires \be {\rm
tr}(Q)=s_u+g_v<0, \s {\rm Det}(Q)=s_u g_v-f_v g_u>0. \ee

The model equations for intracellular calcium waves can be generally
written in a system of reaction-diffusion equations in the form \be
\label{equ-100} \phi_t=D \na^2 \phi +f(\phi), \s \phi=[u \;\; v]^T,
\ee where $D={\rm diag}(D_u, 0)$. The rate $f(\phi)$ is a general
nonlinear function coupling the different components $u$ and $v$.
This nonlinear reaction-diffusion system can be solved using
conventional numerical method or stochastic cellular automata.

\section{Stochastic Cellular Automaton}

Conventionally, reaction-diffusion systems can be solved numerically
using finite difference or finite element methods. Alternatively,
the nonlinear systems can be simulated by using cellular automata
[16]. The cellular automata for reaction-diffusion systems can be
formulated to correspond to the solutions of the related partial
differential equations in a qualitative manner [4] or quantitative
manner [5]. The macroscopic approach via cellular automata as
demonstrated by [16] is always more efficient than explicit
numerical method and can be more efficient than better numerical
technique in many cases.

The discretization of equation (\ref{equ-100}) can be written  as
\be \kk{\phi^{n+1}_{i,j}-\phi^n_{i,j}}{\Delta t}=D
[\kk{\phi^n_{i+1,j}-2 \phi^n_{i,j} +\phi^n_{i-1,j}}{(\Delta x)^2}
 +\kk{\phi^n_{i,j+1}-2 \phi^n_{i,j} +\phi^n_{i,j-1}}{(\Delta
y)^2}]+f(\phi^n_{i,j}). \ee By taking $\Delta t=\Delta x=\Delta
y=1$, this becomes \be \phi^{n+1}_{i,j}= D
[(\phi^n_{i+1,j}+\phi^n_{i-1,j} +\phi^n_{i,j+1}+\phi^n_{i,j-1})-4
\phi^n_{i,j}]+f(\phi^n_{i,j}), \ee which can be written as a generic
form \be \phi^{n+1}_{i,j}= \sum_{k,l=-r}^{r} c_{k,l}
\phi^n_{i+k,j+l}+f(\phi^n_{i,j}), \ee where the summation is over
the $2r+1$ neighborhood. The coefficients $c_k$ are determined from
the discretization of the governing equations, and for this special
case, $c_{-1,0}=c_{+1,0}=c_{0,-1}=c_{0,+1}=D, c_{0,0}=-4D$. In fact,
this is equivalent to a finite-state cellular automaton with a
transition rule $G=[g_{ij}] \;\;(i,j=1,2,...,N)$ from one state
$\Phi^n=[\phi_{ij}^n] \;\;(i,j=1,2,...,N)$ at time level $n$ to a
new state $\Phi^{n+1}=[\phi^{n+1}_{ij}] \;\;(i,j=1,2,...,N)$ at time
level $n+1$. That is, \be G: \Phi^n \mapsto \Phi^{n+1}, \s g_{ij}:
\phi^{n}_{ij} \mapsto \phi^{n+1}_{ij}. \ee The basic assumption for
the rule inference for stochastic automata is that the function \be
g_{ij} (\phi^n_{ij})=\phi^{n+1}_{ij}, \s {\cal V} \le \Gamma
(\phi^n_{ij}, f,r,N), \ee where $\Gamma$ is a function with a range
of $[0,1]$, and ${\cal V}$ is a random variable. At every time step,
a random number ${\cal V}$ is generated for each automaton $(i,j)$.
The new state will only be updated if the generated random number is
greater than $\Gamma$, otherwise, it remains unchanged. Following a
similar derivation for ordinary differential equation [9], the
probability function $\Gamma$ can be written as \be \Gamma
(\phi_{ij}^n,f,r,N)=\Gamma(e^{-f})=a+b e^{-f}, \ee where $a$ and $b$
are coefficients depends on the number of finite states and other
factors such as the accuracy of the approximation to the partial
differential equations. The parameters of the continuum
reaction-diffusion model shall be calibrated to fit the results
given by the stochastic cellular automaton model using a
least-squared procedure so as to get the related accurate transition
rules.

\begin{figure}
\centerline{\includegraphics[width=3in,height=2.5in]{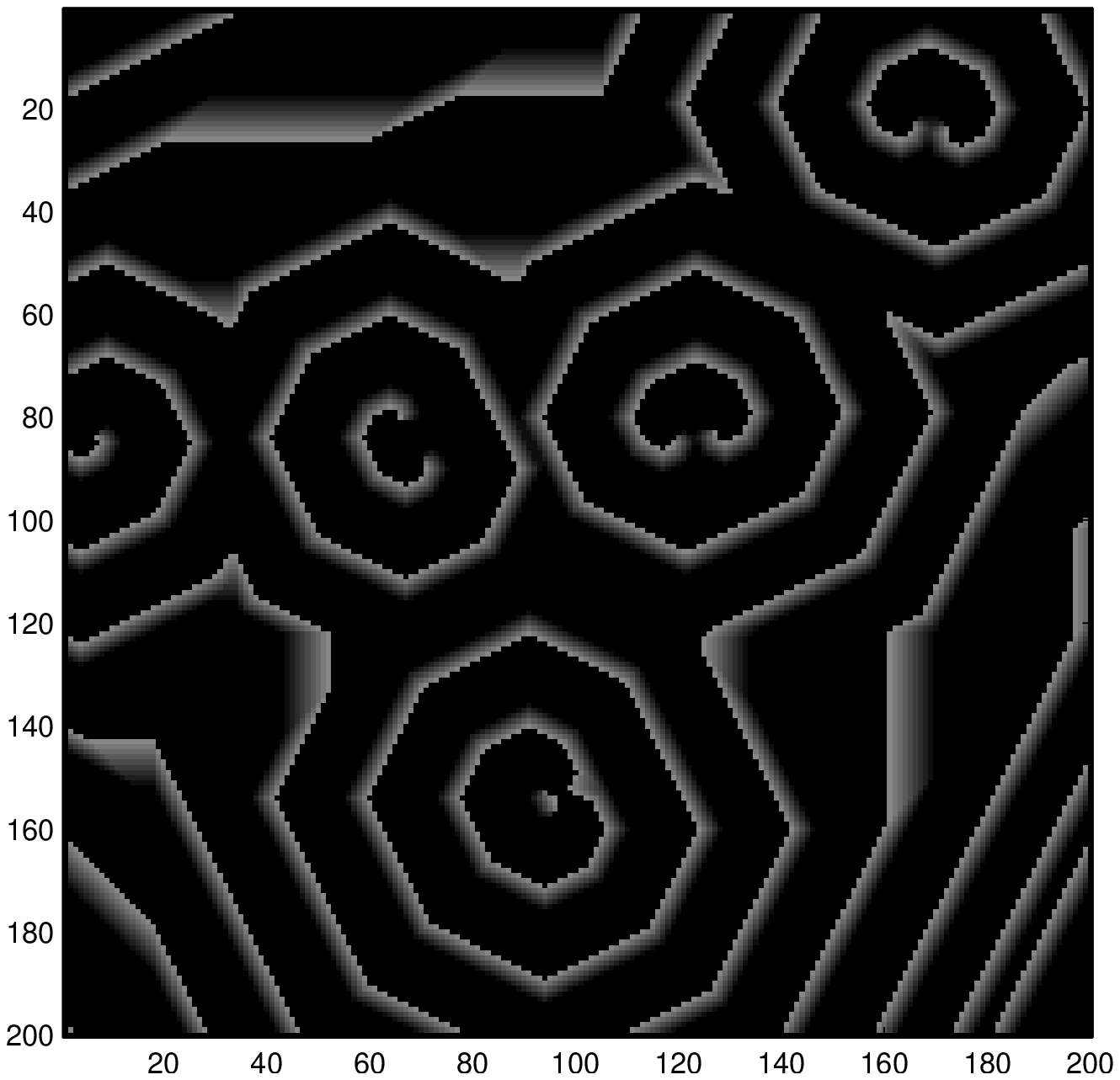}
\includegraphics[width=3in,height=2.5in]{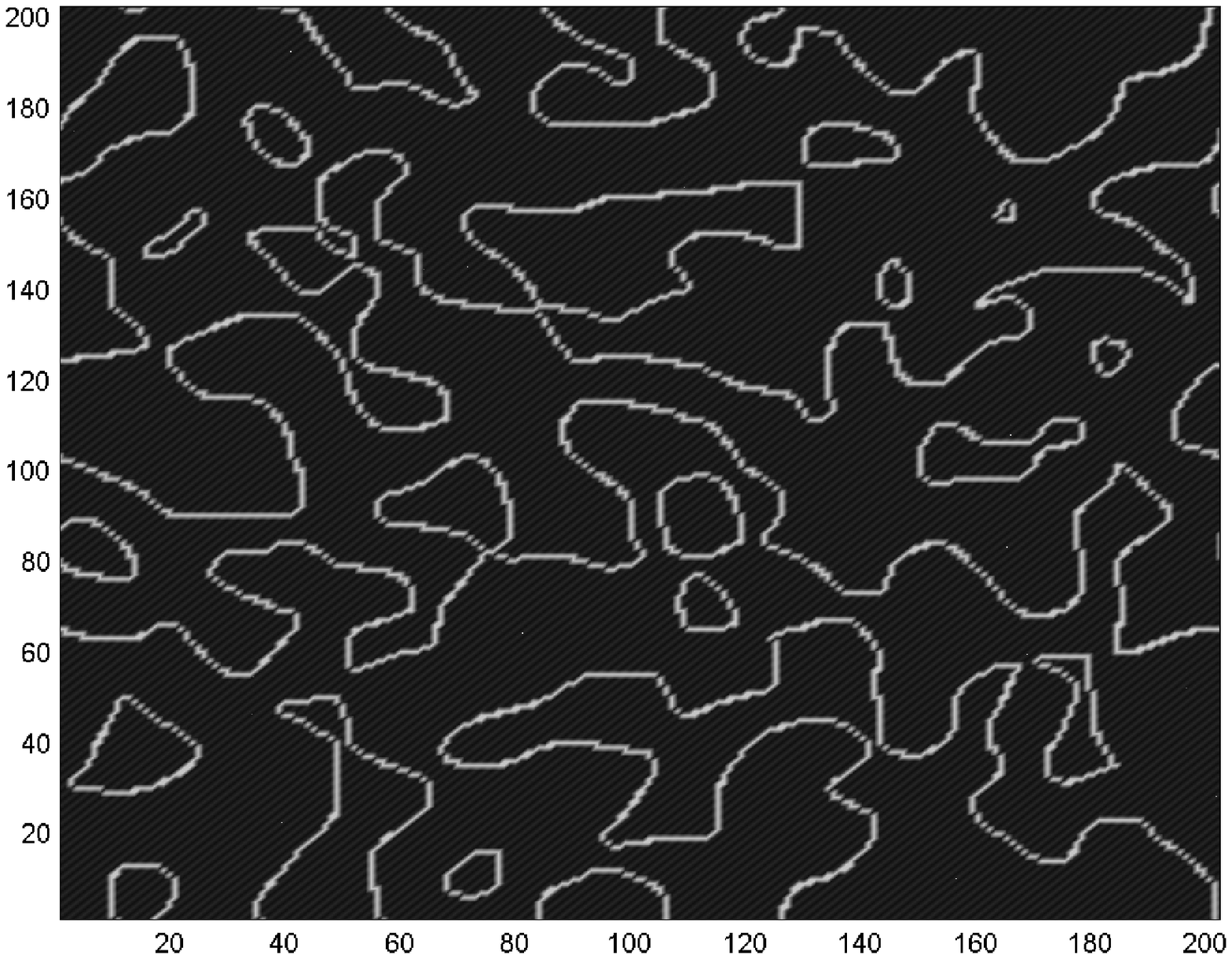}}
\caption{Pattern formation of calcium concentrations (spiral waves
on the left and rings on the right) using values given in section
2.1 }
\end{figure}

\section{Simulations and Results}

By using the approach of stochastic cellular automata with finite
number of states, we can now simulate the reaction-diffusion calcium
transport and nonlinear calcium waves. Numerical simulations are
carried out on an $N \times N$ lattice in 2-D setting, and usually,
$N \ge 40$, or up to $1500$. The number of states is taken to be 256
in the present simulations. Different simulations with different
lattice size are compared to ensure the simulated results are
independent of the lattice size and time steps. In the rest of the
paper, we present some results related to calcium pattern formation,
nonlinear calcium waves and self-organized criticality of the
complexity of reaction-diffusion systems.

\subsection{Pattern Formation and Calcium Waves}

From the initial random configuration, nonlinear reaction-diffusions
can lead to complex patterns. Figure 1 shows two examples of the
calcium concentrations evolving to the interesting patterns in 2-D
configuration with $200 \times 200$ cells. The parameters used in
the calculations are $k=3 \mu M s^{-1}$ (M=Mol/L is the molar
concentration), $k_1=k_3=0.7 \mu M, k_2=0.01 \mu M, k_v=1 \mu M,
k_u=0.27 \mu M, \delta=0.11, \a=0.15 \mu M s^{-1}, \lambda=0.2 s$.
Spiral waves are formed for values $D_u=5 \mu m^2 s^{-1}$ (Fig 1a)
while rings and ribbons are formed for higher value $D_u=25 \mu m^2
s^{-1}$ (Fig 1b). These patterns gradually evolve with time;
however, the general characteristics of patterns only change slowly
with time.

The simulations imply that patterns and structures formed by local
transition rules are relatively stable. Our present results are
consistent with the other well-known studies in pattern formation
such as [11,12] for plants and sea shells by using nonlinear
reaction-diffusion equations. In addition, our present simulations
suggest that patterns arise naturally from the local interactions
either through rule-based/agent-based evolution in terms of
relationships among the neighbourhood individuals in stochastic
cellular automata or partial differential equations in terms of
system variables such as calcium concentrations. The initial
configuration is not important. The rules of interactions or
relationships between entities/individuals in the nearest neighbour
are crucial factors that responsible for the behaviour of the system
and pattern formations.

In addition, the spatial pattern formation of calcium concentration
can provide some positional information of calcium distribution
resulting from the reaction-diffusion transport among cells. This
may have some important implication in the mechanism of calcium
functionality in biological systems. Although many factors such as
the function of proteins, genetic information and enzymes affect the
intracellular transport of calcium, however, the calcium
reaction-diffusion process itself could greatly contribute to the
spatial distribution of calcium and thus be responsible to some
extent for its positional information and signalling in biological
systems. If it is the case, the modelling of calcium transport can
be beneficial to the understanding of formation of the spatial
structure and positional signalling coupled with the genetic and
functional information in biological systems and physiological
mechanisms.

The reaction-diffusion systems of calcium transport are complex.
Nonlinear waves can arise under certain initial and boundary
conditions. For a single source, the calcium concentration starts
to expand and form nonlinear waves as shown in the figure on the
left of Figure 2 which is a snapshot at $t=500$. Numerical
simulations imply  that wave speed decreases with time as expect
for any passive diffusion systems. In addition, the for a random
configuration with periodic boundary conditions, complex nonlinear
wave pattern have observed in numerical simulations as
demonstrated in the figure on the right (Fig. 2b) where calcium
concentration varies with space and time.

\begin{figure}
\centerline{\includegraphics[width=3in,height=2.5in]{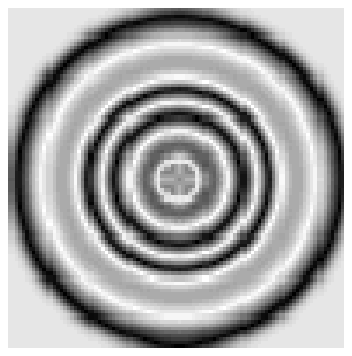}
\includegraphics[width=3in,height=2.5in]{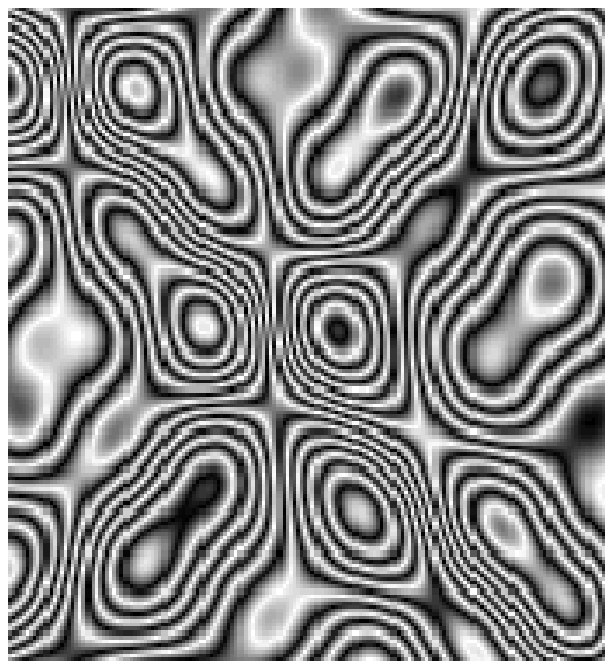}}
\caption{Nonlinear calcium waves for a single source (left) and
multiple sources (right). }
\end{figure}

\subsection{Complexity and Self-Organized Criticality}

For a lattice $N=100 \times 100$ with $256$ states, the complexity
of the cellular automata can be measured by its entropy $S$ is
defined as  $S=-\sum_{i} p_i \log p_i$,  where $p_i$ is the
probability of states $i$ [1]. For a finite state automaton, $p_i$
can be approximated by  the calcium concentration  so that \be
S=-\sum_{i} u_{ij} \log u_{ij}. \ee The variation of complexity or
entropy of the stochastic cellular automata with time is shown in
Figure 3a. It is clearly seen that the complexity varies
significantly at the early stage of the pattern formation process,
then it gradually relaxes to the equilibrium  at long time,
indicating that the reaction-diffusion system is in a quasi-steady
state.

The complex pattern of calcium concentration can be measured by
grouping or counting the number of cells for a given value of
concentration. The results are plotted in Figure 3b. It is clearly
seen that there exists a power law in the distribution of the
number of cells ($n$) versus discrete calcium concentration ($u$).
A least-square fitting of $n \propto u^{-\gamma}$, leads to the
exponent of $\gamma=1.26 \pm 0.02$. This implies the
self-organized criticality in the complex calcium patterns.

\begin{figure}
\centerline{\includegraphics[width=3in,height=2.5in]{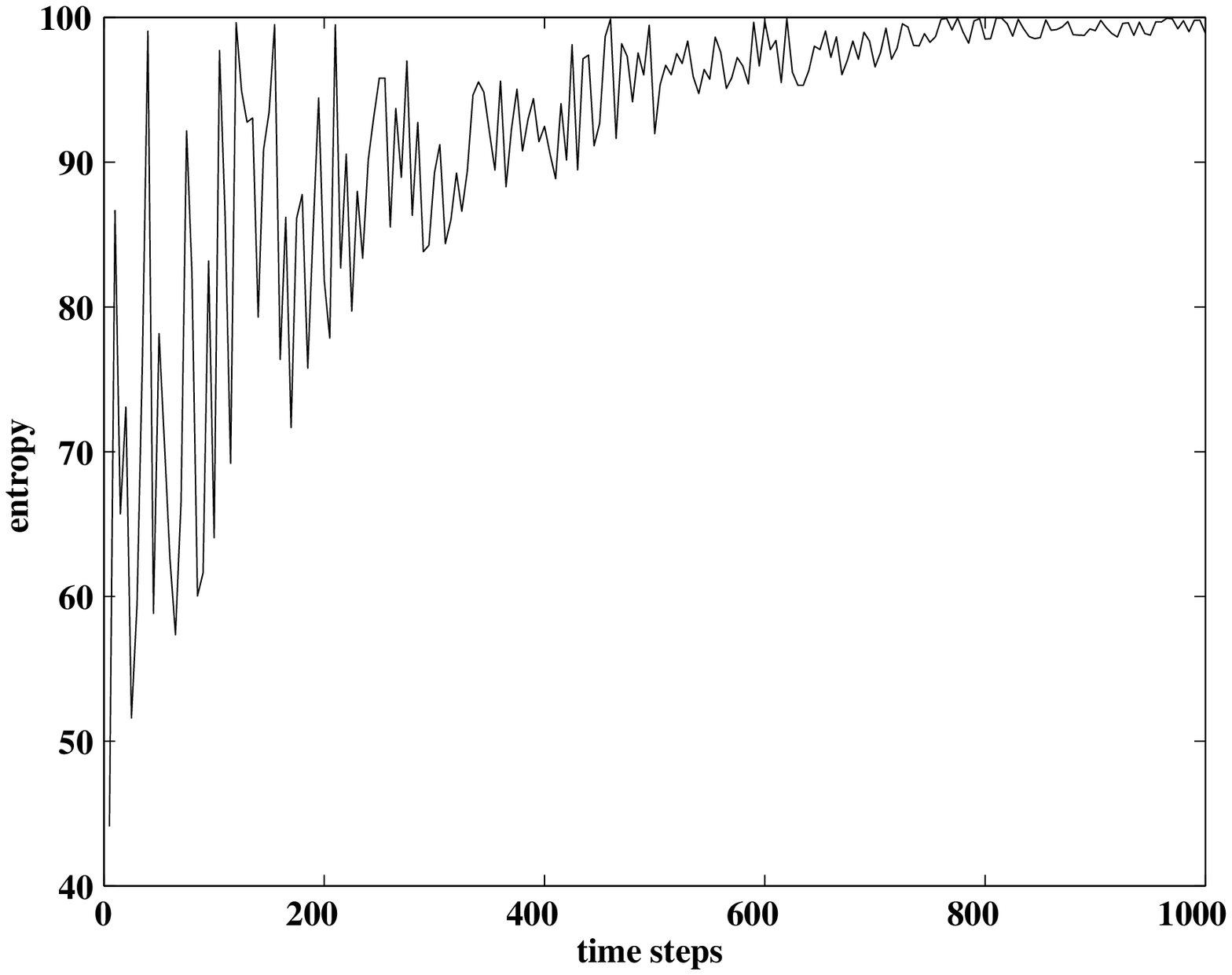}
 \includegraphics[width=3in,height=2.5in]{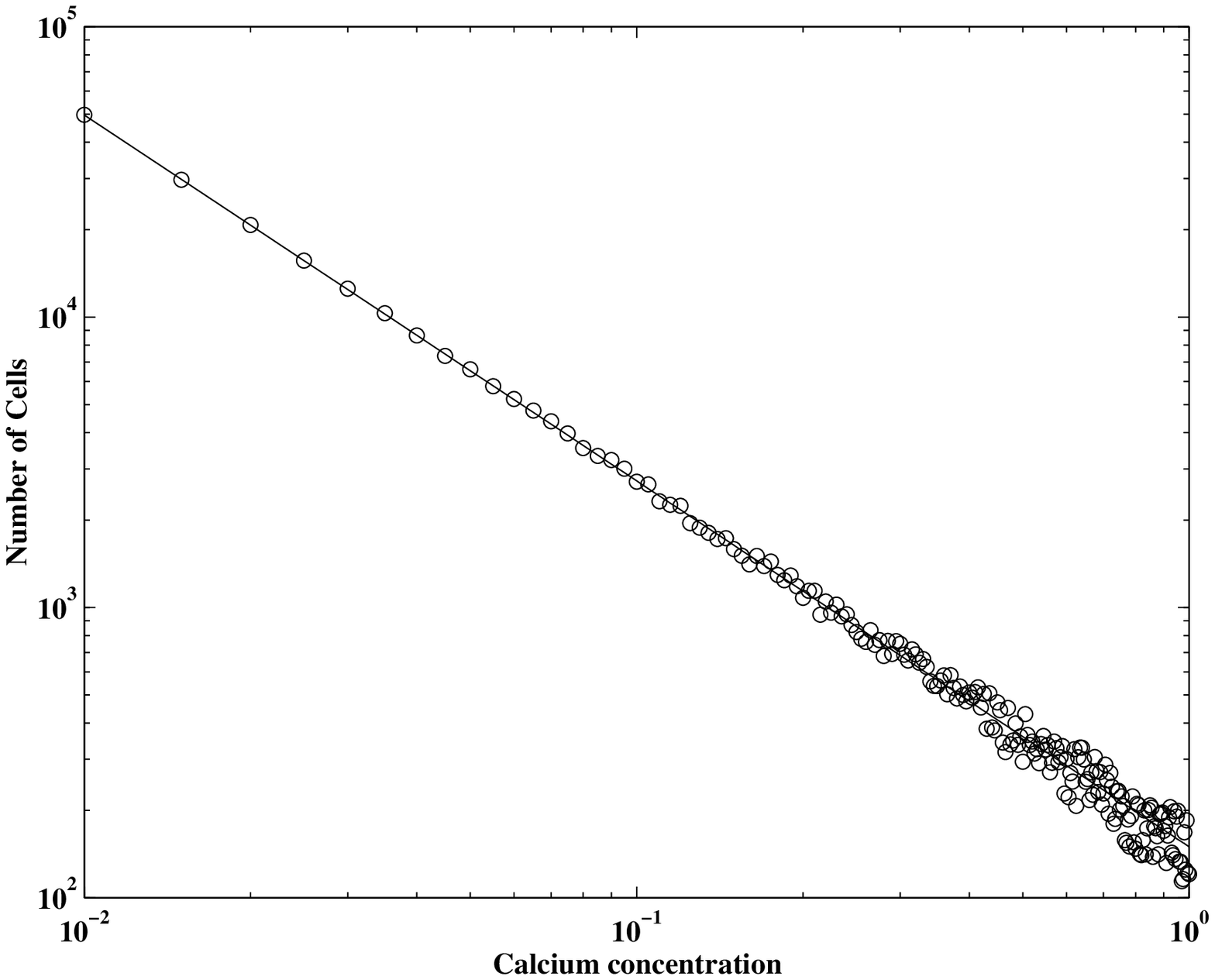}}
\caption{Complexity and entropy variations at different time steps
(left). Self-organized criticality in calcium pattern formation
leads to the exponent $\gamma=1.26 \pm 0.02$ (right).}
\end{figure}
This result may have important implications to the calcium transport
mechanism. Although the detail of intracellular calcium oscillation
and communication is not clear, it is likely that the intracellular
and intracellular interactions mainly occur locally. Thus
reaction-diffusion dominates the process without much contribution
from the convection mechanism. This is consistent with the
physiological aspects of calcium transport and functioning [4,10].

\section{Conclusions}

The finite state stochastic cellular automata have been formulated
to simulate the reaction-diffusion systems of nonlinear calcium
oscillations and waves using the transition rules derived from the
partial differential equations.  By using the proper stochastic
and transition rules between different states, the finite state
automaton can simulate the complexity of calcium transport.
Numerical experiments show that complex patterns can arise from
the initial random configuration due to the local transition rules
between entities of certain nearest neighborhood and the details
of initial configuration is not important. The power-law
relationship between number of cells and calcium concentrations
implies self-organized criticality in the complex patterns.

\section{References}

\begin{description}
\item{[1]} Atri A., Amundson J., Clapham D., and Sneyd J., A
single-pool model for intracellular calcium oscillations and waves
in the Xenopus laevis oocyte, {\it Biophysical Journal}, {\bf 65}
(1993) 1727-1739.

\item{[2]} De Young G. W. and Keizer J., A single pool
IP$_3$-receptor based model for agonist stimulated Ca$^{2+}$
oscillations, {\it Proc. Natl. Acad. Sci. USA}, {\bf 89} (1992)
9895-9899.

 \item{[3]} Dupont G., Goldbeter A., Properties of
intracellular Ca$^{2+}$ waves generated by a model based on
Ca$^{2+}$-induced Ca$^{2+}$ release, {\it Biophysical Journal}, {\bf
67} (1994) 2191-2204.

\item{[4]} Ermentrout G. B., Edelstein-Keshet L.,  Cellular
automata approaches to biological modelling, {\it J Teor. Biol.},
 {\bf 160} (1993) 97-133.

\item{[5]} Gerhardt M., Schuster H., Tyson J.J.,  A cellular automaton model
of excitable media including curvature and dispersion, {\it
Science}, {\bf 247} (1990) 1563-1566.

\item{[6]} Goldbeter A., {\it Biochemical Oscillations and
Cellular Rhythms: the Molecular Bases of Periodic and Chaotic
Behaviour}, Cambridge University Press, Cambridge, (1996).

\item{[7]} Bak P, Tang C and Wiesenfeld K, Self-organized
criticality: an explanation of 1/f noise, {\it Phys. Rev. Lett.},
{\bf 59} (1987) 381-384.

\item{[8]} Bak P, {\it How nature works: the science of
self-organized criticality}, Springer-Verlag, New York, (1996).

\item{[9]} Guinot V.,Modelling using stochastic, finite state
cellular automata: rule inference from continuum model, {\it Appl.
Math. Model.}, {\bf 26} (2002) 701-714.

\item{[10]} Keener J., Sneyd J., {\it Mathematical Physiology},
Springer,  (1998).

\item{[11]} Murray, J. D., {\it Mathematical Biology}, Springer, New York, (1989).

\item{[12]} Meinhardt H., {\it Models of biological pattern formation},
Academic Press, London, (1982).

\item{[13]} Sanderson M. J., Charles A. C., Boitano S., and Dirksen E. R.,
Mechanisms and function of intercellular calcium signaling, {\it
Molecular and Cellular Endocrinology}, {\bf 98} (1994) 173-187.

 \item{[14]} Sneyd J., Charles A. C., and Sanderson M. J.,  A
model for the propagation of intercellular calcium waves, {\it Am J
Physiol.}, {\bf 266} (1994) 293-302.

\item{[15]} Sneyd J., B Wetton, Charles A. C., and Sandeson M. J.,
Intercellular calcium waves mediated by diffusion of ionositol
trisphoshpate: a two-dimensional model, {\it Am. J. Physiology Cell
Physiology}, {\bf 268} (1995) 1537-1545.

\item{[16]} Weimar J. R., Cellular automata for reaction-diffusion systems,
{\it Parallel computing}, {\bf 23} (1997) 1699-1715.

\item{[17]} Wolfram, S., Cellular automata as models of
complexity, {\it Nature}, {\bf 311} (1984) 419-424.

\item{[18]} Wolfram, S., {\it Cellular automata and complexity},
Reading, Mass: Addison-Wesley (1994).

\item{[19]} McKenzie A. and Sneyd J., On the formation and breakup of spiral waves
of calcium, {\it International Journal of Bifurcation and Chaos},
{\bf 8} (1998) 2003-2012.

\item{[20]} Goldbeter A., Depont G., and Berridge M., Minimal model
for signal-induced calcium oscillation and for their frequency
encoding through protein phosphorylation, {\it Proc. Natl. Acad.
Sci. USA}, {\bf 87} (1990) 1461-1465.

\item{[21]} Hunding A. and Ipsen M., Simulations of waves in calcium
models with 3D spherical geometry, {\it Mathematical Biosciences},
{\bf 182} (2003) 45-66.

\item{[22]} Romeo M. M. and Jones C. K. R. T., The stability of travelling
calcium pulses in a pnacreatic acinar cell, {\it Physica D}, {\bf
177} (2003) 242-258.

\item{[23]} Wilkins M. and Sneyd J., Intercellular spiral waves of
calcium, {\it J. Theor. Biol.}, {\bf 191} (1998) 299-308.

\item{[24]} Barkley D., Linear stability analysis of rotating spiral
waves in excitable media, {\it Phys. Rev. Lett.}, {\bf 68} (1992)
2090-2093.

\item{[25]} Sneyd J., Falcke M, Dufour F. F., Fox C., A comparison
of three models of the inositol trisphosphate recptor, {\it Prog.
Biophys. Mol. Biol.}, {\bf 85} (2004) 121-140.

\item{[26]} Falcke M, Reading the patterns in living cells -- the
physics of Ca$^{2+}$ signaling, {\it Adv. Phys.}, {\bf 53} (2004)
255-440.

\end{description}
\end{document}